\documentclass[twocolumn,showpacs,amsmath,amssymb]{revtex4-1}
\usepackage{graphicx} 
\usepackage{bm} 
\usepackage{epsfig}

\begin{document}

\title{ Quantifying the sorting efficiency of self-propelled run-and-tumble swimmers by geometrical ratchets}

\author{I. Berdakin$^{1,2}$, A. V. Silhanek$^3$, H. N. Moyano$^1$, V. I. Marconi$^{1,2}$, and C. A. Condat$^{1,2}$}

\affiliation{$^1$Facultad de Matem\'atica, Astronom{\'i}a y F{\'i}sica, Universidad Nacional de C\'ordoba, X5000HUA C\'ordoba, Argentina.\\
     $^2$IFEG-CONICET,\\ X5000HUA C\'ordoba, Argentina.\\
     $^3$D\'epartement de Physique, Universit\'e de Li\`ege, B-4000 Sart Tilman, Belgium.\\}


\begin{abstract}

Suitable asymmetric microstructures can be used to control the direction of motion in microorganism populations. This rectification process makes it possible to accumulate swimmers in a region of space or to sort different swimmers.  Here we study numerically how the separation process depends on the specific motility strategies of the microorganisms involved. Crucial properties such as the separation efficiency and the separation time for two bacterial strains are precisely defined and evaluated. In particular, the sorting of two bacterial populations inoculated in a box consisting of a series of chambers separated by columns of asymmetric obstacles is investigated. We show how the sorting efficiency is enhanced by these obstacles and conclude that this kind of sorting can be efficiently used even when the involved populations differ only in one aspect of their swimming strategy.  

\end{abstract}



\maketitle


\section{Introduction}

Self-propelled objects moving in confining environments at low Reynolds numbers exhibit interesting physical properties, some of which are not yet well understood and deserve to be studied in view of their technological applications. These objects range from artificial microswimmers that can be actuated upon by using applied magnetic fields~\cite{keaveny2013} to motile cancer~\cite{mahmud2009, konstant2013} and stem~\cite{peng2011} cells, to motile bacteria~\cite{galajda2007, kim2010} and spermatozoa~\cite{elgeti2010, denis2012}. The study of their properties in confined regions has been made possible by recent advances in microfabrication~\cite{squires2005, leung2012} and observation~\cite{wilson2011, altshuler2013} techniques.

Aside from the intrinsic problems posed by the motion of confined, self-propelled, run-and-tumble microorganisms, there is strong interest in the biomedical and engineering communities in efficiently controlling microorganism motion. A physical, nondestructive method of achieving this control is suggested by the geometrically-induced guidance caused by the walls of asymmetric openings, the funnels. This ratchet effect was first used by Galajda and coworkers to generate  an inhomogeneous bacterial distribution~\cite{galajda2007}. The otherwise random motion of bacteria was controlled, {\it i.e} rectified, by the funnels in the box. Since bacteria swim parallel to the funnel walls, it is easier for them to cross the barrier from the wide to the narrow funnel opening than in the opposite direction. The funnels then define a preferred direction for the swimmer motion, leading to an increase in the cell concentration on one side of the box and a consequent decrease on the other.  This effect was also recently  observed in the puller eukaryote swimmer {\em Chlamydomonas reinhardtii }~\cite{kantsler2013}. Interestingly, the rectification process can be reversed if chemotactic or collective motions prevail, as shown in Refs.~\cite{Lambert2010,
drocco2012}. Various aspects of the microorganism dynamics  have been the subject of recent studies~\cite{condat2005, peruani2007, garcia2011, disalvo2012,  gregoire2004, bertin2009, Weber2013, Romanczuk}.

This rectification effect can be particularly useful when there are mixtures of microorganisms exhibiting different motility strategies. In this connection it is worth mentioning that various microfluidic techniques for sorting motile microscopic objects have been developed in the last few years. This is the case for the separation of motile from non-motile sperm cells~\cite{cho2003}, the sorting of {\em E. coli} by length~\cite{hulme2008}, the use of self-driven artificial microswimmers for the separation of binary mixtures of colloids~\cite{yang2012} and the study of the dynamics of several kinds of particles combining asymmetric obstacles and a time-dependent voltage~\cite{bogunovic2012}. One of their objectives is to eliminate the cellular stress and damage associated with alternative techniques such as centrifugation. Geometrical sorting also avoids the use of applied fields or chemical gradients, whose maintenance at scales of the millimeter or longer is difficult~\cite{voldman2009}. Would it be also possible to use the rectification effect to efficiently sort cells by their swimming strategies? This is the question we would like to answer in this paper.

Given their ubiquity, motile bacteria are of particular interest. They swim by rotating thin helical filaments called flagella; each flagellum is driven by a rotary motor powered by the flow of ions ($H^+$ or $Na^+$) across the cytoplasmic membrane. In order to take advantage of chemotactic gradients, many of these bacteria have evolved a run-and-tumble swimming strategy~\cite{Berg1993}. In the case of the paradigmatic {\em Escherichia coli}, during the run mode the flagella rotate counterclockwise and the microorganism moves in a forward, relatively straight direction, whereas during the tumble mode, one or more flagella rotate clockwise and the bacterium is reoriented in a new direction~\cite{Berg2004}. As shown forty years ago by Berg and Brown for {\em E. coli}~\cite{Berg1972}: (A) runs are not strictly straight due to rotational diffusion, (B) run lengths are exponentially distributed, and (C) bacterial heading changes at tumbles by less than $90 ^\circ$, preserving some memory of the previous run, a fact that is often neglected in theoretical treatments. It is worth noting that some marine bacteria show anti-persistency, in what is called a run-and-reverse strategy~\cite{barbara2003,xie2011}.

 The case of bacterial directed motion under asymmetrical geometrical confinement, first observed and explained in Ref.~\cite{galajda2007}, was modeled phenomenologically in Ref.~\cite{wan2008}. The authors considered point-like swimming bacteria following run-and-tumble dynamics with a constant motor force magnitude and thermal fluctuations. Although this model neglects the details of the swimmer dynamics, it reproduces the most important experimental findings and has been an inspiration for further theoretical work. In Ref.~\cite{tailleur2009}  the relation between the ratchet effect and symmetry breaking by the funnel array geometry was clarified. It was shown there that the break of time-reversal symmetry needed for rectification is provided by the forced rotation of bacteria when colliding with a wall, and not by the motor force of bacteria. This is so because the break of time-reversal symmetry provided by the bacterial motor is lost at a coarse-grained level of diffusion where detailed balance is restored. Later, the influence of the specific dynamical properties, from (A) to (C), described by Berg and Brown, on the accumulation of cells in presence of asymmetric obstacles was studied in detail in Ref.~\cite{berdakin2013}. This numerical analysis used experimental values of the motility parameters. It was found that different swimming strategies may yield very different microorganism accumulation efficiencies, being measured as the device capacity to concentrate cells (number of concentrated cells/number of inoculated cells of the same type). We summarize the main results of that work:

\begin{enumerate}
     \item 	In unbounded environments there are two processes that degrade the orientational correlation: tumbling and rotational diffusion. The first is much more important for systems with short runs, while rotational diffusion gives the dominant contribution to memory loss in systems  characterized by long runs. These effects can be quantified by the velocity correlation function.
     \item 	A study of the mean square displacement in unbounded environments reveals that the translational diffusion coefficient $D_{T}$ decreases strongly as the change-of-heading angle at a tumble increases. Unless typical runs are very long, $D_{T}$ is approximately given by its value in the absence of rotational diffusion. These results indicate that, to make accurate predictions about swimmer sorting, it is necessary to consider the specific motility properties of the microorganisms involved.
     \item 	When the dynamics of free swimmers is incorporated into a spatially constrained environment (asymmetric geometry) long run lengths and small tumble emergence angles lead to an increased cell density near the walls and, consequently, to fast net displacement in the easy ratchet direction. In general, long permanence near the walls and suitable wall-of-funnels architecture, i.e., funnel walls at least as long as the run length and funnel openings of the order of the cell size, favor rectification or cell concentration.
     \item 	Increasing the average run duration, $\tau$, increases the time of permanence close to the walls, leading to enhanced rectification. If $\tau$ is negligible, the swimmers cannot take advantage of swimming along the funnel walls and directed motion does not ensue.
     \item 	Increasing the average speed during the run and decreasing the average change-of-heading angle at a tumble, i.e. increasing persistence, enhance wall accumulation and rectification. 
     \item 	Good agreement was obtained with available experimental data, specifically regarding the time of rectification and the efficiency of a microfabricated wall of
funnel-shaped openings as the one used in Ref.~\cite{galajda2007}.
     
 \end{enumerate}

In this work, we use the improved  phenomenological model introduced by Berdakin {\it et. al.} in Ref.~\cite{berdakin2013} to investigate the efficiency of asymmetric microarrays used as sorting devices, and their dependence
 on the swimming strategies of the microorganisms involved. Our objective is to help to design good sorters, using a model that incorporates  real motility parameters. In Section 2 we review the computational model and define the quantities to be calculated, such as the extraction time and the sorting efficiency. In Section 3 we present our numerical results, which are briefly discussed in the concluding section. 

\section{Methods}

\begin{table}
\caption{\label{tab:tab1} Motility parameters of two different {\em E. coli} strains: $s_1$
corresponds to AW405 and $s_2$ to CheC497 in Ref.~\cite{Berg1972}.}
\begin{tabular}{ccccccc}
\hline
Swimmer & $\bar{v} [\mu m/s]$ & $\sigma_{v} [\mu m/s]$ & $\bar{\phi} [^\circ]$ & $\sigma_{\phi} [^\circ]$ & $\tau [s]$ & $D_R [rad^2/s]$ \\
\hline\hline
$s_1$ (wild type)   & 14.2  & 3.4 & 68 & 36 & 0.86 & 0.18 \\
$s_2$ (mutant)   & 20.0  & 4.9 & 33 & 15 & 6.30 & 0.06 \\
\hline
\end{tabular}
\end{table}

\begin{figure*}[ht]
\begin{center}
 \includegraphics[angle=0, width=17.78cm]{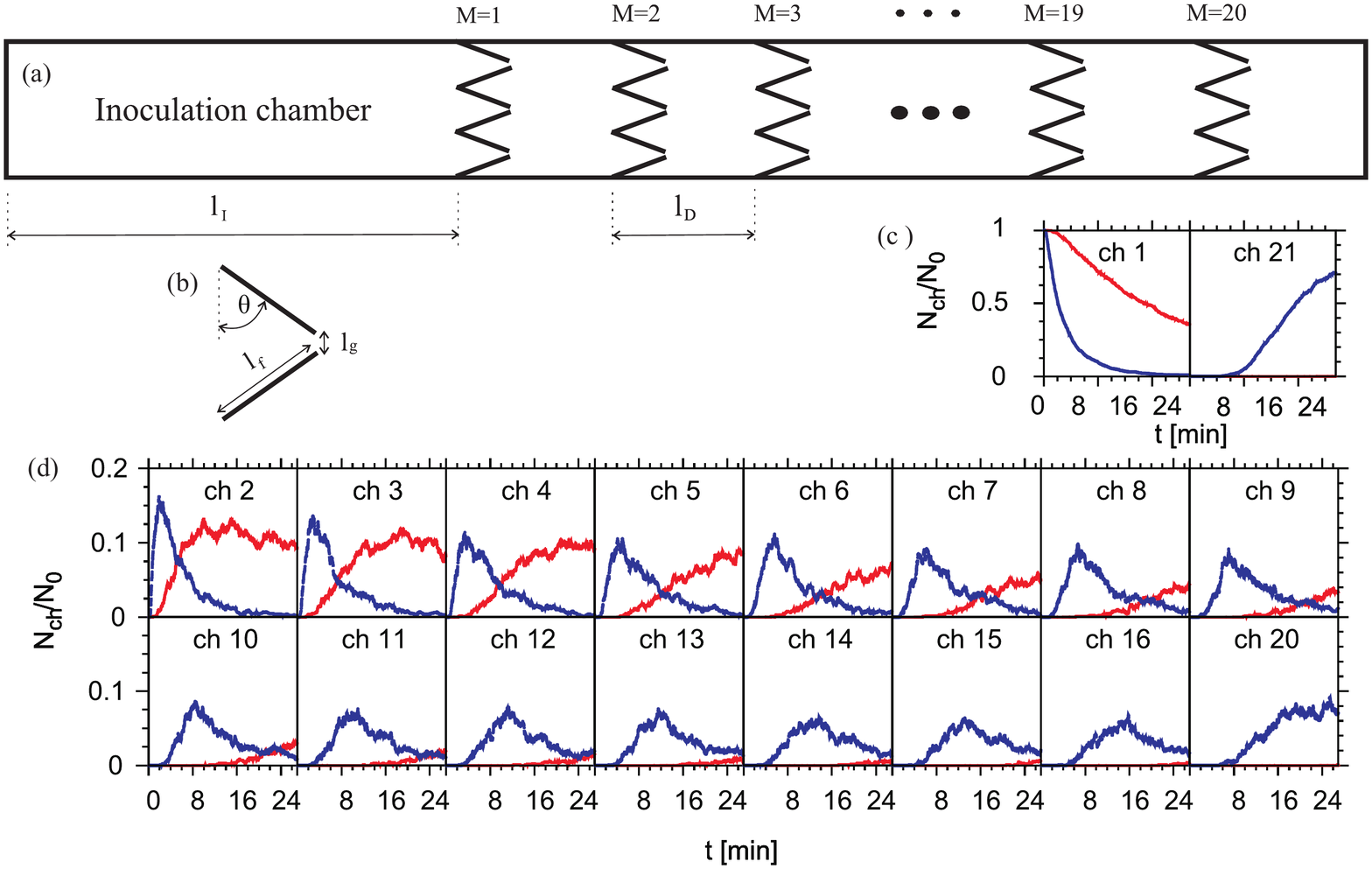}
\caption{Color online. (a) Sketch of the geometry of  a $21-$chamber microarray with $M = 20$ identical asymmetric funnel columns and $N_f=3$ (number of funnels). All chambers, except for the first inoculation chamber, $l_I$, have the same length $l_D = 150 \mu m$. (b) Single funnel geometry showing its relevant parameters: $l_f$, $l_g$ and $\theta$. (c) Time evolution in minutes of the normalized bacterial populations ($N_{ch}/N_0$) in the inoculation (leftmost) and last (rightmost) chambers where $N_0$ is the number of cells of each kind inoculated at $t = 0$ $s$.  The bacterial populations are wild-type {\em E. coli} and a mutant used in~\cite{Berg1972}: $s_1$ (red/grey) and $s_2$ (blue/black) respectively. Only $s_2$ is present in the last chamber in the window of time considered, $26$ min. The  time to pick up  $s_2$ completely pure in $ch 21$  is $t^* = 24.1 $ min and the $s_2$ extraction efficiency is  $\epsilon_{\%} = 85$ $ \%$.  (d) Time evolution as in (c), but  in the intermediate chambers.  Chambers 17 to 19 are not shown but are all similar, without $s_1$.}
\label{fig:sch}
\end{center}
\end{figure*}

{\bf The model.} We study numerically a dilute system of $2N_0$ microscopic self-propelled particles, \emph{the swimmers}, moving under low Reynolds number conditions and confined to a micro-patterned two-dimensional box of size ${ L_{x}{\times} L_{y}}$. The box contains $M$ identical, equidistant columns of obstacles, each consisting of $N_f$ openings (asymmetric funnels), of gap size $l_g$ and wall length $l_f$ (see Fig.~\ref{fig:sch}(a)). We choose $N_f = 3$ and $M$ from 5 to 20. The $M-1$ inner chambers and the last chamber have all the same length $l_{D}=150 $ $\mu m$. The length of the inoculation (leftmost) chamber is kept constant, $l_I=450$ $\mu m$, in order to have a fixed fractional occupied area to define a high dilution initial condition at $t=0$. The relevant geometrical parameters are illustrated in Fig.~\ref{fig:sch}: (a) for the array configuration and (b) for the  single-funnel shape.

Each swimmer, whose location is given by a vector $\vec {r}_{i}$, is represented by a soft disk of radius $r_{s}$, moving in two dimensions with speed $v_{i}$ and heading in the direction of the unit vector $\hat {\bf v}(\Phi_{i})=\cos(\Phi_{i}) \hat{\bf i} + \sin(\Phi_{i} )\hat{\bf j}$. In a confined space, the swimmer dynamics is determined by the overdamped equation of motion,

\begin{equation}
		\gamma\vec{v}_i = \vec{F}_i^m + \vec{F}_i^{sw} + \vec{F}_i^{s}
		\label{eq:v}
\end{equation}

where $\gamma$ is the medium damping constant and the acting forces are explained in detail 
below. The runs described by Eq.~(\ref{eq:v}) are interrupted by tumbles and affected by rotational 
diffusion, all of which results in a change of swimmer heading given by,

\begin{equation}
		\Delta \Phi_i = \Delta \phi \chi + \nu\sqrt{2D_R\Delta t} (1-\chi)
		\label{eq:phi}
\end{equation}

where $\nu$ is a Gaussian-distributed random number, $\chi$ is a state variable equal to $0$ during a run and $1$ during a tumble, $D_R$ is the rotational diffusion coefficient and $\Delta t$ the numerical integration time step.
Aside from the frictional damping, the forces operating on the microswimmer are:

\begin{enumerate}
     \item      {\em Self propulsion.} When starting to move under low Reynolds number conditions, swimmers in an unbounded fluid get almost instantly to a constant final speed. In our model this constant speed is given by ${F}^m/\gamma$, 
     where 
      ${F}^m$ is the modulus of the propelling force. The initial condition for the swimmer population speed is chosen from a  
 		normal distribution  with mean $\bar{v}$ and standard deviation $\sigma_v$. Based on this distribution, each cell is assigned a given speed at $t=0$, which remains the same during the whole simulation.
 		
 The heading of the swimmer is altered only by tumbling or rotational diffusion. Tumbles are assumed to be 	instantaneous (real  tumbles last  $0.1$ seconds in mean about $10$ $\%$ of the run time for wild type \emph{E. coli} and a smaller percentage for longer run bacteria - see Table~\ref{tab:tab1} ). Tumbles result in a rotation, $\Delta\phi$,  from the previous direction of motion, which  we consider Gaussian-distributed and centered at $\bar{\phi}$  with a width $\sigma_{\phi}$, (see Table~\ref{tab:tab1} and Ref.~\cite{berdakin2013}). Successive  tumbles are 
		spaced by almost straight runs  exponentially distributed with mean duration $\tau$. During a run, asymmetries in the self propulsion system and environmental
		fluctuations result in deviations from a perfectly straight path. These  deviations are measured by the 
		rotational diffusion coefficient, $D_R$, and included in our model via the changes in the swimmer heading~\cite{li2009} expressed by Eq.~(\ref{eq:phi}). 

     \item	 {\em Interaction with the walls.}  It is modeled by a steric repulsive force $\vec{F}_{i}^{sw} $ normal to the walls,
     
\begin{equation}
		\vec{F}_i^{sw} = f^{sw}(1-r_{ik}/a)\Theta(1-r_{ik}/a)\hat{n}_k, 
		\label{eq:fsw}
\end{equation}

		where $f^{sw}$ is the maximum strength of the force, $\Theta$ is the step function, $\hat{n}_k$ is a unit vector normal to the {\em k-th} wall, $r_{ik}$ is the distance between the
		{\em i-th} particle and the center of the {\em k-th} wall, $a = r_s + w/2$, and $w$ is the wall width.
	        Since the swimming direction is unchanged during the collision  and
	        the normal component of the propulsion force is counteracted by the repulsion of the wall, the swimmer keeps lightly bouncing against the wall. The component of  $\vec{F}^m$ that is parallel to the wall propels the bacterium along the wall with a reduced speed that is proportional to the sine of the angle formed by the incidence direction and  $\hat{n}_k$.  This phenomenological representation of the interaction has the interesting property of reducing the speed of the cells when they swim parallel to a wall without the need of adding an extra parameter to the model. The speed reduction of a bacterium swimming parallel to a wall has been studied experimentally in Ref.~\cite{Frymier1995}. Either a tumble or rotational diffusion may allow the swimmer to move away from the wall.

	         This interaction is responsible for the observed accumulation at the walls~\cite{li2009} and for the directed motion and sorting of bacteria~\cite{galajda2007}. As remarked with measurements of wall accumulation  for bacteria with different swimming strategies
		in Ref.~\cite{mino2011}, the wild-type {\em E. coli} was significantly less attracted to the surfaces than a mutant strain that does not display tumbling.   

     \item     	 A purely steric {\em swimmer-swimmer repulsion} of maximum intensity $f^s$,

	\begin{equation}
           \vec{F}_{ij}^{s} = f^{s}(1-\mid\vec{r}_{ij}\mid/2r_s)\Theta(1-\mid\vec{r}_{ik}\mid/2r_s)\vec{r}_{ij},
        \label{eq:fss}
        \end{equation}	

		with $\vec{r}_{ij} = \vec{r}_i - \vec{r}_j$.  The hydrodynamic interaction between microswimmers is not important at very low swimmer concentrations~\cite{li2009}, and we disregard it here. Our approximations are buttressed by recent measurements of cell-cell and cell-wall interactions using {\em E. coli}, which show that thermal and intrinsic stochasticity wash out the effects of long-range fluid dynamics~\cite{drescher2011}. These experimental results imply that physical interactions between bacteria are mainly determined by steric collisions and lubrication forces.

 \end{enumerate}

A comparison with the model of reference \cite{wan2008} is in order. In that reference the runs were 
assumed to have a constant duration, all the swimmers moved with the same speed and 
started in a completely random direction after each tumble (the emergence angles are 
uniformly distributed in [$0,2\pi$]). Our model differs from that of Ref.~\cite{wan2008} in all these aspects. 
A further difference is that we take into account rotational diffusion, which was neglected in 
Ref.~\cite{wan2008}, where instead the center-of-mass motion is affected by thermal random forces. 
These changes were already introduced in Ref.~\cite{berdakin2013} to obtain a more faithful description of 
the observational facts.
  
Taking into account all the interactions described above we arrive to our set of dynamical equations to be solved numerically~\cite{berdakin2013} for the $2N_0$ run-and-tumble microswimmers.  We assume that  the mixed swimmers population  is initially randomly distributed  in the inoculation chamber.
Using a fourth order Runge-Kutta  algorithm we integrate the dynamical equations of motion and we obtain the  trajectories  for each confined swimmer. The  averages over realizations are later performed.    
 For simplicity we will always compare only two swimmer strategies, a situation that is easiest to implement in the laboratory using two different fluorescent markers. Of course more than two swimmers could also be sorted as shown in Ref.~\cite{berdakin2013}.
Table~\ref{tab:tab1} specifies the motility parameters of the swimmers simulated in this work, the wild type \emph{E. coli}, $s_1$, and a faster, less frequently tumbling mutant, $s_2$. The radius of the soft disks is taken to be $r_{s}=0.5$ $ \mu m$ for all swimmers.
The optimal single-funnel geometric parameters  for an efficient rectification of wild-type {\em E. coli}, $s_1$, were found in Ref.~\cite{berdakin2013} to be $l_g = 2$ $ \mu m$, $l_f = 30$ $ \mu m$, and $\theta = 68^\circ$, so we keep these parameters fixed for all simulations. The box width used here is $L_y = 80$ $ \mu m$ and the wall width is taken to be $w = 2$ $ \mu m$ for both, the box walls and the funnel walls. The width of the inoculation chamber, $l_I = 450$ $ \mu m$, has been chosen to keep an initially low swimmer density. The number of swimmers, $N_0$, of each strain is adapted to maintain an initial occupied area fraction  of $0.05$ at the inoculation chamber for all array geometries. If  $\gamma = 6\pi\eta r_s$, the frictional drag coefficient, $r_s = 0.5$ $\mu m$ and $\eta = 10^{-2}$ poise (the viscosity of water at $20^{\circ}C$), then $\gamma \sim 9.425$ x $10^{-6}$ $g/s$.
Under these conditions, the strength of the motor force of a bacterium swimming at 20 $\mu m/s$ is $0.17$ $pN$. The magnitudes of the forces acting upon the swimmers are $f^s = 200$ and $f^{sw} = 300$ in units of $\gamma$, equivalent, respectively, to ten and fifteen times the force exerted by the motor on the fluid at 20 $\mu m/s$. With this choice in our phenomenological model  bacteria penetrate no more than 10 $\%$ of $r_s$ inside walls or other bacteria.

{\bf Calculated quantities.}
With the aim of quantifying the efficiency of the sorters we propose two parameters as convenient indicators of the separation process: (a) the {\em separation time}, $t^*$, defined as the time elapsed between the arrival of the first swimmer in the fast, $s_2$, class, and that of the first swimmer in the slow, $s_1$, class, to the last  chamber (chamber from where a pure cell population could be extracted or concentrated), and  (b) the {\em separation efficiency}, $\epsilon_{\%} $, which we define as the fraction of the fastest type that has arrived at the last chamber by the time $t^*$.   It is convenient to define the percent extraction efficiency as follows, 
\begin{equation}
\epsilon_{\%} = 100\frac{N_F(t^*)}{N_0},
\label{eq:ef}
\end{equation}
 being
$N_F(t^*)$ being the number of swimmers of the fast species that is present in the last chamber at $t^*$, when its 
purity is still $100 \%$.

\section{Results}

\begin{figure}
\begin{center}
 \includegraphics[angle=0, width=9cm]{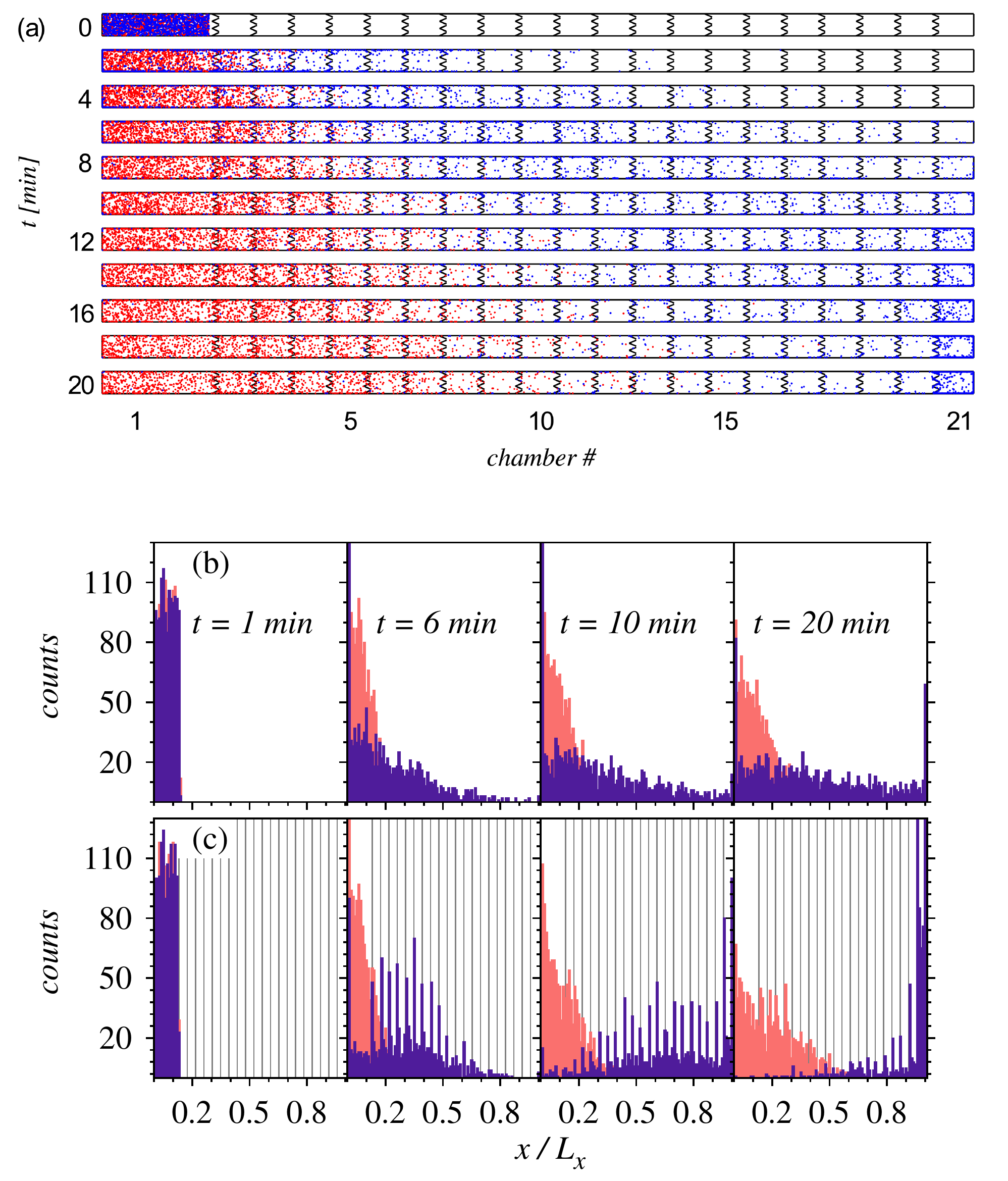}
\caption{Color online. (a) Snapshots illustrating the separation of the two bacterial types considered in Fig.~\ref{fig:sch}. After being uniformly inoculated in the first chamber, both populations are rapidly segregated. (b) Comparative profiles for the spatial evolution of the $s_1$ (red/grey) and $s_2$ (blue/black) bacterial populations for a clean box (upper histograms) and for the 21-chamber box of the same overall size (lower histograms) at the indicated times.  The 21-chamber box is a far more efficient extraction device than the single-chamber  box. 
}
\label{fig:snap}
\end{center}
\end{figure}

We first consider a sample with $M = 20$ funnel columns and two homogeneous bacterial distributions  initially inoculated in the first chamber. 
These bacteria are wild-type {\em E. coli} and a mutant studied by Berg and Brown~\cite{Berg1972}; their characteristic dynamical parameters are specified in Table~\ref{tab:tab1}. We compute the variations of the total bacterial populations of each mutant in the first and the last chambers, which are shown in Fig.~\ref{fig:sch}(c). After $30$ independent realizations, the average separation time for this system is $t^*= 16$ min and the average separation efficiency for the mutant $s_2$ is $\epsilon_{\%} = 67 \pm 19\%$.
Three factors contribute to the high extraction efficiency for this mutant: its higher average speed, its higher persistence, i.e. low $\bar{\phi}$, and, mainly, the longer duration of the runs, which increases both $D_T$ and the contact time with the rectifying walls. The advance of both populations through the various chambers is shown in Fig.~\ref{fig:sch}(d), where we see that the purification process improves with successive chambers. From chamber 12 onwards, we also observe that the time evolution of the $s_2$ pulse (blue) is almost position-independent until it reaches the last column. 
Instantaneous snapshots of the bacterial populations considered in Fig.~\ref{fig:sch} are shown, as functions of time,  in Fig.~\ref{fig:snap}(a), where they are seen to start from a uniform distribution in the inoculation chamber and advance at different rates in the easy ratchet direction. A comparison between corresponding panels in Figs.~\ref{fig:snap}(b) and \ref{fig:snap}(c) shows clearly how these rates are enhanced by the ratchet geometry of the column array, giving an estimated $5$ $ \mu m/s$ drift velocity for the $s_2$ population, five times larger than that found for $s_1$.  As a result of  these different velocities inside the box, both populations are soon largely separated and can be readily sorted out.

It is interesting to compare what happens in the specially designed box, an array of funnel columns,  with the result obtained in a single channel  with  the same area and clean of obstacles,  when the bacterial populations are subject to the same initial conditions. In the clean box, as the histograms in Fig.~\ref{fig:snap}(b) show, the “fast” type also moves forward first, in part taking advantage of its longer runs along the side walls, but the separation is much less efficient than for the funnel-containing box, for which at  $t = 20$ min there is no $s_1$ swimmer from chamber 16 to 21. Purification is complete there. For the particular realization represented in Fig.~\ref{fig:snap}(b) the extraction time and extraction efficiency are, respectively,  $t^* = 24.1$ min and $\epsilon_{\%} = 85\%$.  At very long times, a uniform distribution is expected for the single-chamber configuration, while an exponentially increasing population of each bacterial type is expected in the specially designed box. This exponential increase is responsible for the high concentrations near the end of the array of columns, which permit the extraction of a high fraction of the first bacterial type arriving there. This situation is clearly shown in Fig.~\ref{fig:snap}(c), where it is also possible to observe $s_2$ concentration spikes where the obstacle columns are located. The stronger tendency of $s_2$ to concentrate near the walls, as compared to $s_1$, was recently studied in detail~\cite{berdakin2013}.  

\begin{figure}[ht]
\begin{center}
\includegraphics[angle=0, width=8.5cm]{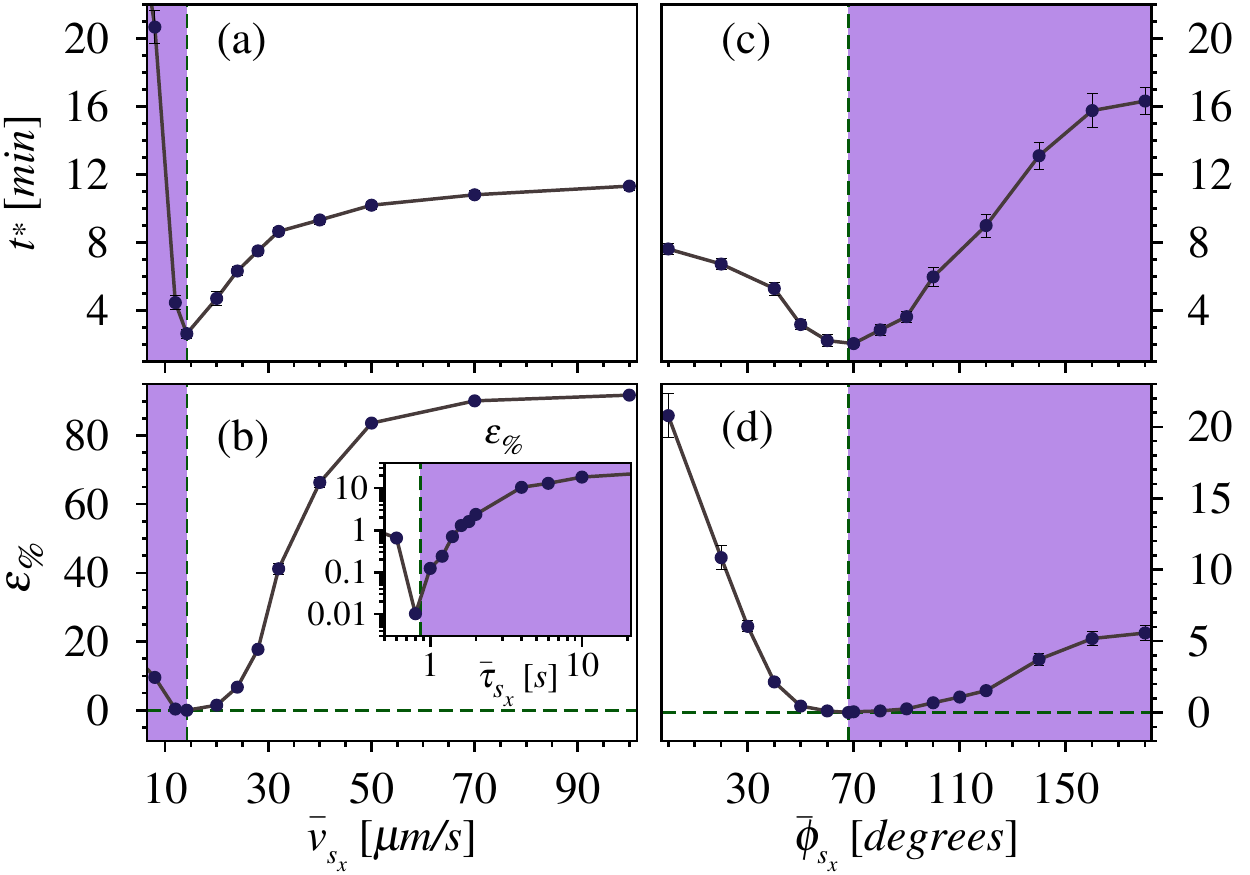}
\caption { Color online. (a) Extraction or pick up time and (b) sorting efficiency for the fastest swimmer to reach the last chamber when we simultaneously simulate wild-type {\em E. coli}, $s_1$, and a mutant, $s_x$, for which only the average run speed is changed. (c) and (d): the same quantities when only the average tumbling angle  $\bar {\phi}_x $ of the mutant is changed. Inset: sorting efficiency when only the mean run duration, $\tau_x $, of the second swimmer is changed. Note the different vertical scales between (b) and (d). Here  we use a smaller array with $M=10$.} 
 \label{fig:par}
\end{center}
\end{figure}

Now we study the sorter efficiency of two swimmers (one real and the second real or artificial) as a function of the 
specific dynamical parameters characterizing the microswimmers. 
In Fig.~\ref{fig:par}  we show the extraction times and sorting efficiencies for two swimmers, one of which is wild-type {\em E. coli}, $s_1$, and the other  $s_x$,  for which a single motility parameter is changed. From Figs.~\ref{fig:par}(a) and \ref{fig:par}(b), for which only the mutant speed was changed, we see that, when $v_x < v_1 = 14.2$ $ \mu m/s$ (shadowed region), the wild-type bacteria arrive first and can be purified during a time $t^*$. This purification window is, for instance, of $20.3$ min if $v_x = 8$ $ \mu m/s$. The window gets narrower when $v_x$ is close to $v_1$, but grows monotonically when $v_x > v_1$. Similarly, the sorting efficiency has a minimum when $v_x = v_1$ but increases with the difference between bacterial speeds. We can purify $18\%$ of $30$ $ \mu m/s$ mutants and we have 7.5 minutes to do it. This behavior was to be expected, since a faster bacterium diffuses farther, and more importantly, can take advantage of longer runs parallel to the walls.   The saturation value of $t^*$ is given by the average time it would take the ``slow'' bacterial strain to travel from the inoculation chamber to the end of the box (this would be the separation time for a hypothetical infinitely fast strain). The inset in panel (b) shows that varying the mean run duration has an effect qualitatively similar to changing the mean speed ($\tau_1 = 0.86$ $ s$). Although changing the mean tumbling angle, see panel (c), yields extraction times of the same order as changing the mean speed, the corresponding sorting efficiencies, panel (d), are markedly lower. In this case, we compared  a hypothetical   swimmer $s_x$, whose average tumbling angle $\bar {\phi}_x $  is modified, with the wild type, for which $\bar {\phi}_1  = 68^\circ $.  Easiest to separate are the persistent-walk bacteria, for which $\bar {\phi}_x = 0^\circ $ and the run-and-reverse bacteria, for which $\bar {\phi}_x  = 180^\circ $. It is worth noting that {\em large $t^*$ does not necessarily mean large $\epsilon_{\%}$}.  For example, if  $\bar {\phi}_x  = 180^\circ $, $t^* = 16$ min, but $\epsilon_{\%}$ is only $5\%$, a relatively low value when compared with $\epsilon_{\%} = 20\%$ that results for  $\bar {\phi}_x = 0^\circ $, for which $t^*$  is only $7$ min.

\section{Discussion}

We have investigated  arrays of asymmetric-funnel columns built for the purpose of concentrating or sorting out one type of self-propelled swimmer in a run-and-tumble microorganism mixture. As characteristic parameters to measure the suitability of a given architecture, we introduced the extraction time and the sorting efficiency. The first is important because it gives us the length of the temporal window available to the experimentalist to pick up the chosen strain, but does not tell us anything about the number of swimmers ready to be extracted. This is given by the separation or sorting efficiency.   

The separation efficiency depends both on the motility parameters of the swimmers and on the geometrical dimensions of the device, which we can modify according to the swimmer types we are dealing with. Here we have considered the competition between swimmers having different intrinsic dynamical properties. Currently,  we are working out in detail the effect of modifications in the geometrical array parameters that define the asymmetric confining system. 

The following are some predictions from our study:

\begin{itemize}
\item	Asymmetric funnel arrays are capable of sorting diluted distributions of run-and-tumble swimmers {\em in a controlled way}, enhancing the efficiency obtained using a box free of geometrical constraints.
\item	A sizable fraction of the chosen swimmers can be $100\%$ purified   even if the original mixture is composed of swimmers that are dynamically {\em only slightly different}.
\item	In general, unless the motility properties of the swimmers are very similar, for $M$ of the order of 10 the extraction time should be long enough to allow the experimentalist to purify the sample.
\end{itemize}

In our simulations we did not include fluid flow, so that the net bacterial motion from left 
to right is solely due to funnel asymmetry. Under flow our results would be very different.  
Flow in a narrow channel is known to change the accumulation of cells on the walls and 
even to cause upstream swimming. Moreover, the response to flow depends upon the 
tumbling rate of the cells~\cite{altshuler2013,constanzo2012}. We could hypothesize that flow may lower the sorting efficiency when 
pointing in the easy ratchet direction (left-right) and enhance the efficiency otherwise, but 
this is something that deserves careful study.

In this paper our goal was to efficiently sort swimmers at low concentrations as experiments in view of how technological applications in this field are generally made.  But what would happen at high 
concentrations? One way to look at these problems is to adapt the well-known Vicsek's 
model~\cite{vicsek1995,greogoire2003, bertin2006, chate2008}.
Hydrodynamic equations have also been obtained in the high density limit using a 
Boltzmann approach \cite{bertin2006} and through the coarse-graining of the microscopic 
dynamics \cite{Ihle2011}. Recently, Drocco and coworkers \cite{drocco2012} added steric repulsion 
to the Vicsek flocking algorithm and studied the motion of self-propelled particles in a 
confining microenvironment such as the one considered in this paper. These authors 
found rectification effects induced by the high particle concentration in the absence of 
preferential motion along the walls. The nature of this rectification process is therefore 
quite different from the one we have considered here and opens the way to the analysis of a 
possibly rich phenomenology and other types of applications. 

Another extension of the studies in this paper that would be specially profitable in the case of 
the smallest self-propelling microorganisms could be made by explicitly considering the 
influence of passive and active fluctuations on the system behavior.  
 The impact of the different fluctuation types on the collective dynamics 
of active Brownian particles with velocity alignment has already been studied \cite{grossmann2012, romanczuk2012}.
  Our understanding of microswimmer dynamics would be enhanced by 
the analysis of the behavior of these active particle systems in asymmetric confining 
microarchitectures.  

To summarize, the purpose of this paper was twofold: First, to introduce new definitions, 
those of extraction efficiency and of separation time, which are advantageous to quantify 
how effective is a given microarchitecture to sort different types of run-and-tumble self-propelling microorganisms. Second, to specifically investigate how these microorganisms 
can be sorted by their motility strategies. We have shown how testable predictions can 
be made using {\em realistic bacterial parameter values.}  These predictions can be very useful to design efficient microfluidic devices. We further point out that, although
run-and-tumble strategies are common in the bacterial world, this type of motion is not restricted to bacteria. The locomotion of the unicellular alga {\em chlamydomonas} exhibits, in the dark, nearly straight swimming runs interrupted by abrupt changes in direction. The run distributions are exponentially distributed, with $\bar{\tau} = 11.2$ $ s$~\cite{Goldstein2009}. Consequently, the dynamics of this eukaryote are likely to be describable by the model discussed in this paper  as well.
Our numerical calculations can be easily generalized to include the possibility of bacterial birth/death during the experiment, work we have in progress. 

{\bf Acknowledgments}

This work was supported by CONICET (PIP 112-200801-00772 and PIP 112-201101-00213) and SeCyT-UNC (Projects No. 05/B354 and No. 30720110101526) (Argentina), FNRS (Belgium)-CONICET,  FWO-MINCyT FW/09/04 and KU Leuven-UNC bilateral projects.

\end{document}